\documentclass[3p, onecolumn]{elsarticle}
\usepackage{amssymb}
\usepackage{graphicx}
\newcommand{{\ddtphi}}{\ddot\phi}
\newcommand{{\dtphi}}{\dot\phi}
\newcommand{{\ud}}{\mathrm{d}}
\newcommand{{\la}}{\langle}
\newcommand{{\ra}}{\rangle}

\newcommand{\mpl}{M_\mathrm{pl}}

\newcommand{{\mpc}}{\,\mathrm{Mpc}}
\newcommand{\vs}{\vec{s}}
\newcommand{\bs}{\bar{s}}
\newcommand{\hs}{\hat{s}}
\newcommand{\vb}{\vec{b}}

\newcommand{\da}{\Delta}
\newcommand{\Ip}{{I^\prime}}
\newcommand{\vdelta}{\vec{\delta}}
\newcommand{\vv}{\vec{v}}
\newcommand{\pd}{\partial}
\newcommand{\valpha}{\vec{\alpha}}
\newcommand{\vbeta}{\vec{\beta}}
\newcommand{\vphi}{\vec{\phi}}
\newcommand{\veff}{V_{\mathrm{eff}}}

\begin{document}

\title{A Meandering Inflaton}

\author[cnl]{S.-H. Henry Tye} 
\ead{sht5@cornell.edu}
\author[cnl]{Jiajun Xu}
\ead{jx33@cornell.edu}
\address[cnl]{Laboratory for Elementary Particle Physics \\ Cornell University, Ithaca, NY 14853, USA}
\date{\today}

\begin{abstract}
If the cosmological inflationary scenario took place in the cosmic landscape in string theory, the inflaton, the scalar mode responsible for inflation, would have meandered in a complicated multi-dimensional potential. We show that this meandering property naturally leads to many e-folds of inflation, a necessary condition for a successful inflationary scenario. This behavior also leads to fluctuations in the primordial power spectrum of the cosmic microwave background radiation, which may be detected in a near future cosmic variance limited experiment like PLANCK. 
\end{abstract}


\maketitle

\section{Introduction} 

It is generally believed that there is a cosmic landscape in string theory \cite{KKLT}. 
One expects the string scale to be in between the Hubble and the Planck scale, $H \ll m_s \ll \mpl$, so that a single Planck scale field range in the landscape naturally contains many string scale features. When the vacuum energy is not too far below the string scale, the universe is expected to undergo cosmic inflation, which describes how our universe began. The inflationary universe scenario is strongly supported by theoretical considerations and observational data. 
For our purpose here, the cosmic landscape may be approximated by a $d$-dimensional random potential, and the inflaton (the scalar mode for inflation) path appears smooth in the coarse-grained limit, but is actually rugged with fine-grained string scale features. In this paper, we will present two generic properties of such an inflationary scenario:
\begin{enumerate}
\item At least 60 e-folds of inflation is necessary in an inflationary scenario to solve the well-konwn cosmological problems such as the flatness and the horizon problems. To achieve this in the slow-roll inflationary scenario usually requires an almost flat smooth potential. Without any fundamental theory, one can always invent such a potential. However, in the context of string theory, such an almost flat potential requires some fine-tuning. While a generic potential may not be flat enough, it is not necessarily smooth either. A rugged $d$-dimensional potential is more realistic in the context of the cosmic landscape (Fig.\ref{fig_pot}), where $d \gg 1$. For large enough $d$, percolation probability approahes unity so the wave function of the universe is mobile in the landscape \cite{Tye:2006tg,Sarangi:2007jb}. For a Planck scale field range, one expects many string scale features and consequently the inflaton takes a meandering path. Since the de-Sitter quantum fluctuation ($H$) is smaller than the size of the features ($m_s$), the turns of the inflaton path may be treated as classical scatterings. We will prove that an arbitrary detour of the inflaton path always increases the travel time, leading to more e-folds than given by the corresponding coarse-grained smooth potential. With appropriately many features in the potential, enough e-folds of inflation becomes generic. Although the scenario is motivated by the cosmic landscape, if a specific string compactification is complicated enough, a brane motion inside the bulk can exhibit similar meandering behavior.

\item The above scenario naturally leads to fluctuations in the the cosmic microwave background radiation (CMB) temperature  (TT) and polarization  (TE/EE) power spectra. These fluctuations are due to the turns and detours of the inflaton path in the $d$-dimensinal potential.
Due to the randomness of the features in the potential, such fluctuations appear with irregular spacings, magnitudes and shapes \cite{Tye:2008ef}. 
While some of these features are probably too fine and/or small to be detected, a few big fluctuations may already have been observed in the  data \cite{Nolta:2008ih,Nagata:2008tk, Nagata:2008zj,Nicholson:2009pi}. 
With more data coming in the future, one expects this prediction to be well tested. The existence of such fluctuations will have a profound impact on the inflationary scenario, on the origin of our universe and on the existence and properties of the cosmic landscape.   
\end{enumerate}

\section{Detours Always Lead to More E-folds}

Let $\vec{\phi} \equiv (\phi_1, \phi_2, \dots, \phi_d)$ be the canonical inflaton field in the $d$-dimensional field space (indices are raised by $\delta^{IJ}$), and let $V(\vphi)$ be the actual $d$ dimensional scalar potential. We assume that over relatively large field distances, the potential appears coarse-grained smooth, denoted by $V_c(\vphi)$, which may or may not be flat enough to support 60 e-folds of inflation, and may suffer from the $\eta$ problem. For a small enough field range, we are safe to keep only the linear term in the Taylor-expansion of $V_c(\vphi)$. 
Without loss of generality, for a specific small field range, we can rotate to the basis $\vec{\phi} = \{\phi_\Ip, \sigma\}$ ($\Ip$ denotes all directions transverse to $\sigma$.), so that $V_c(\vphi)$ for this segment has gradient only along the $\sigma$ field,
\begin{equation}
\label{bgV}
V_c(\vec{\phi}) = V_0 - b\, \sigma ~, \quad 0 \le \sigma \le \sigma_f ~. 
\end{equation}
where the constant $b>0$ and $\sigma_f$ denotes the validity range of the linear expansion. This yields a straight inflaton path from 
$\vec{\phi}={\bf 0}$ to $(0,\dots, \sigma_f)$. When the inflaton moves beyond $\sigma_f$, one needs to re-expand the potential with a new slope $b'$ and a new $\sigma$ direction. In this way, one can approximate an arbitrary coarse-grained smooth potential using a set of piecewise linear segments.

In this work, we take the Hubble parameter $H \approx \mathrm{const}$ and we use the slow-roll approximation ($3H\dot{\phi_I}= - \nabla_I V$) for the field range $[0, \sigma_f]$. The qualitative properties are robust even when these approximations are relaxed.  

\begin{figure}[h]
\begin{center}
\includegraphics[width=7cm]{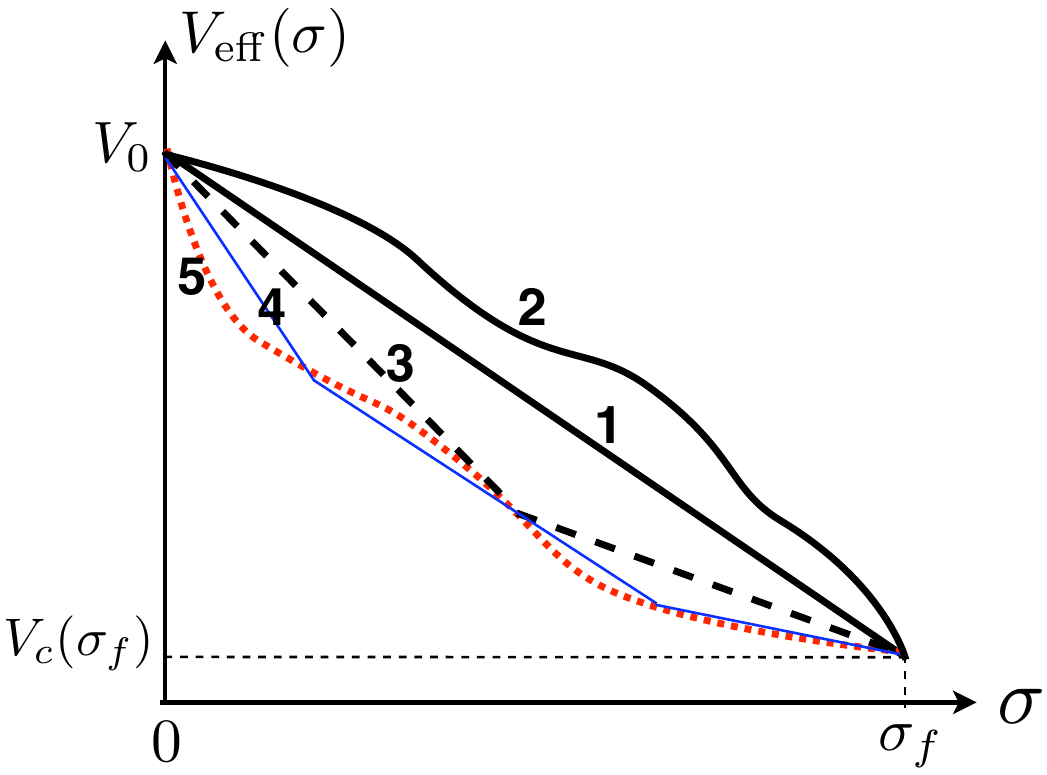}
\includegraphics[width=8cm]{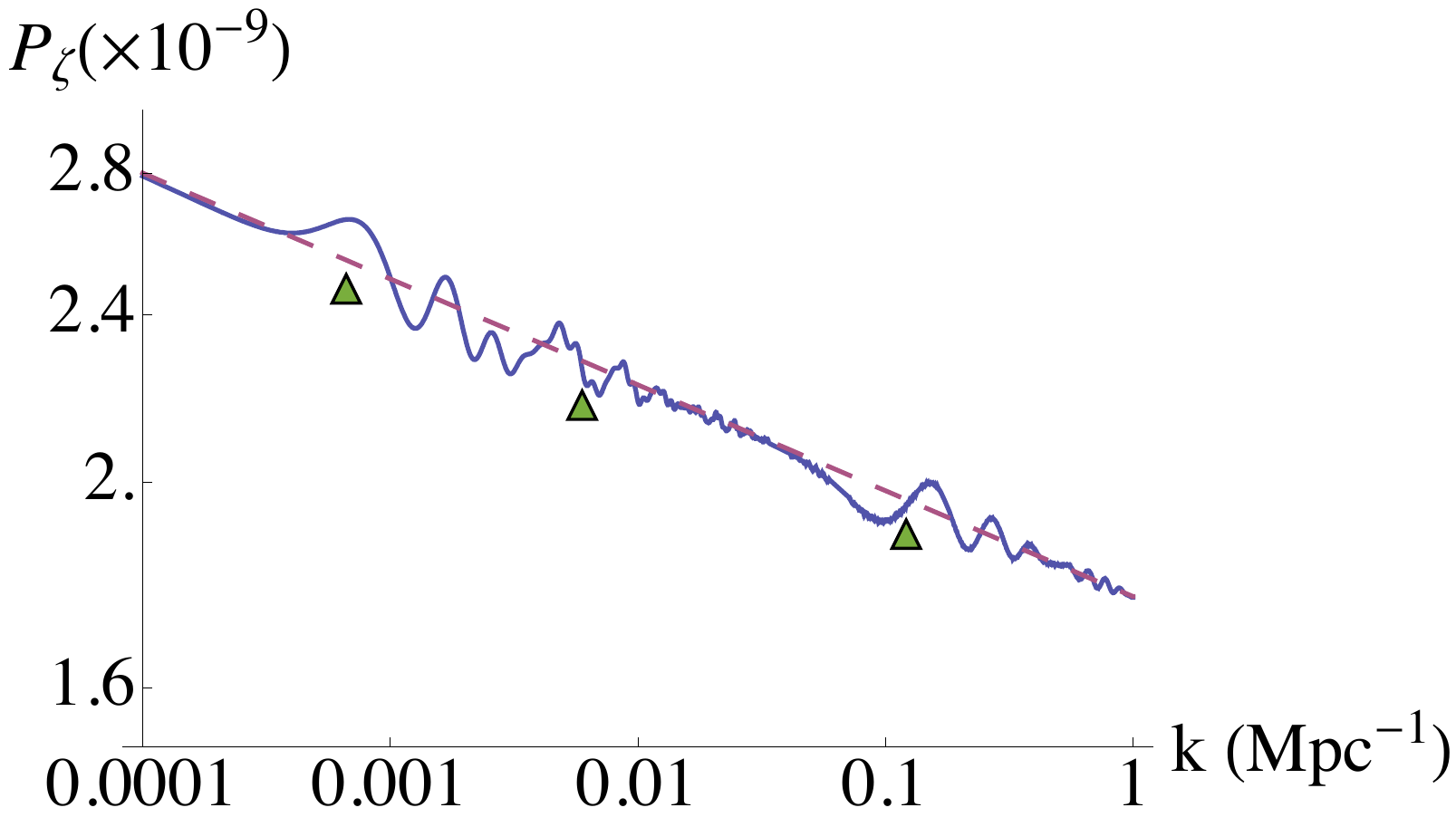}
\end{center}
\caption{\label{fig_pot} Left Panel. Consider a potential segment that begins at $\sigma=0$ at $V_0$ and ends at $\sigma_f$ at $V_c(\sigma_f)$, each curve above can be regarded as either the \emph{actual}  potential $V(\phi)$ for $d=1$ or the \emph{effective} potential $\veff(\sigma)$ for $d>1$. For $d=1$, in the slow-roll approximation, the linear segment $^\#$1 yields the smallest number of e-folds. Suppose $^\#5$ (red dotted curve) is the actual potential, then $^\#1$, $^\#3$ ($K=3$) and $^\#4$ (blue solid thin line, with $K=4$) are successively finer approximations of potential $^\#5$, yielding increasing number of e-folds, $N_e(^\#5) > N_e(^\#4) > N_e(^\#3) > N_e(^\#1)$. 
For $d>1$, small transverse motions always increase $N_e$. That is, $\veff(\sigma)$ always yields more e-folds than the corresponding  potential shown in the figure.}
\caption{\label{fig_pk} Right Panel. We show the smooth power spectrum with $n_s=0.95$ (dashed line), and an illustration of the power spectrum with three features labeled by the triangles. We choose the magnitudes of the features $B^i \sim 0.01$. For single field $d=1$, the features correspond to $\sim 1\%$ sudden change in the potential gradient. However, for multifield $d \gg 1$, the features arise due to the turning of inflaton trajectory. The turning angle can be quite large $\Delta_{IJ} \sim 1$, and still be compatible with data since $B^i \sim \Delta_{IJ}/d$ for the multifield case. Although features arise due to different reasons in the single field and multifield scenario, one can not tell the difference from their appearance in the power spectrum.}
\end{figure}

Now the actual potential is rugged instead of smooth. Let us introduce fine grained features to $V_c(\vphi)$ (\ref{bgV}) to reproduce or to better approximate the actual potential. Starting with $\vec \phi ={\bf 0}$ at $t=t_1=0$, the slope $\vec{s} \equiv (s_1, s_2, \dots , s_d)$ changes at $t_k$ from $\vs(k-1)$ to $\vs(k)$ ($t_{k+1} > t_k ~, k=1,\dots, K-1$),
\begin{equation}\label{sk}
s_I(k) =  b \, \delta_{dI} + \sum^{k}_{j=1}  \delta_I(j) 
\end{equation}
where $\vdelta (j)$ is a constant vector parameter for the $j$-th step. The potential for $t_{k} \le t < t_{k+1}$ now becomes (denote $\vec{\phi}(j) \equiv \vec{\phi}(t_j) $)
\begin{equation}
\label{radpot}
V (\vec{\phi})= V_0 - \sum_{j=1}^{k-1} \vs(j) \cdot \left[ \vec{\phi}(j+1) - \vec{\phi}(j) \right] 
- \vec{s}(k) \cdot \left[ \vec{\phi} - \vec{\phi}(k) \right] ~.
\end{equation}
Since we are treating $V_c(\vphi)$ as the background potential, we require that, after $K$ steps, $\vphi$ will reach $\vphi(K) \equiv (0,\dots, \sigma_f)$, i.e.
\begin{eqnarray}
\phi_\Ip(K) &=& \sum_{j=1}^{K-1} \frac{s_\Ip (j)}{3H} \, ( t_{j+1} - t_j ) = 0 ~, \label{i_cstr}\\
\sigma(K) &=& \frac{b \,t_K}{3H} + \sum_{j=1}^{K-1} \frac{s_\sigma(j)}{3H} \, ( t_{j+1} - t_j ) = \sigma_f ~. 
\label{d_cstr}
\end{eqnarray}
We also require that both the coarse-grained smooth potential $V_c(\phi)$ (\ref{bgV}) and the actual potential $V(\phi)$ (\ref{radpot}) begin at $\vec{\phi} = {\bf 0}$ with value $V_0$ and end at $\vec{\phi} = (0,\dots, 0,\sigma_f)$ with value $V_c(\sigma_f)$ (as illustrated in Fig.\ref{fig_pot}), namely
\begin{eqnarray}
\label{V_cstr}
\sum^{K-1}_{j=1} \sum_{\Ip, \sigma} s_I (j) \left[ \phi^I (j+1) - \phi^I (j) \right] = b\, \sigma_f ~.
\end{eqnarray}
Let us count the number of free parameters. For $K$ steps, we introduce $K-1$ time parameters $t_k$ ($2 \le k \le K$). For each $\phi_\Ip$ and $\sigma$, we introduce $K-1$ fluctuation parameters $\delta_I(i)$, altogether $(K-1)d$ parameters. Taking into account the $d+1$ constraints from Eq.(\ref{i_cstr}), Eq.(\ref{d_cstr}) and Eq.(\ref{V_cstr}), the number of parameters is $(K-1) + (K-1)d - (d+1) = (K-2)(d+1)$. The actual potential can be approached in the $K \to \infty$ limit. 

To proceed, we need to properly choose $\sigma_f$, so that not only the linear expansion is valid, but also the $\sigma$ field is monotonic and can be used as our ``clock''. (Otherwise, we have to choose a finer-grained piece-wise linear approximation.) As we shall see, this also limits the sizes of the kinks. We now show that \emph{an arbitrary detour from $V_c(\vphi)$ (\ref{bgV}) will take longer to reach $V_c(\sigma_f)$}. Such an increase in the number of e-folds of inflation is unbounded from above. 

Using $\sigma$ as the variable, we can rewrite the $d$-dimensional potential $V(\vphi)$ as an effective 1-dimensional potential $\veff (\sigma)$. For $d=1$, $\veff(\sigma)$ reduces to $V(\phi)$. 
\begin{eqnarray}
\label{radpot1}
&& \veff (\sigma) \;=\; V_0 - b \,\sigma -  \hat{S}(k) \left[ \sigma - \sigma (k) \right] 
- \sum_{j=1}^{k-1} \hat{S}(j) \left[ \sigma (j+1) - \sigma(j) \right] ~, \\
&& \sum_{j=1}^{K-1} \hat{S}(j)[\sigma(j+1)-\sigma(j)] \;=\; 0 ~, \quad \hat{S}(j) \;=\;  \hs_{\sigma}(j) + \sum_\Ip \frac{\hs^2_\Ip(j)}{b + \hs_\sigma(j)} ~, \label{VScstr}
\end{eqnarray}
The last equation is simply the constraint (\ref{V_cstr}). In this sense, Fig.\ref{fig_pot} may also represent the $\veff (\sigma)$ for $d>1$.

We now show how arbitrary features of the potential increases the number of e-folds when compared to the smooth linear potential. Let us also generalize the piecewise constants $\hs_I$ and $\hat{S}$ to continuous functions of $\sigma$ (one may view this as a $K \rightarrow \infty$ limit) so the number of e-folds can be expressed as
$N_e =  \int \ud t H = \int \ud \sigma\, 3H^2/(b + \hs_\sigma(\sigma))$. 
Here $\hs(\sigma) \equiv \vs(\sigma) - \vb$, i.e., $\hs_\sigma = s_\sigma - b$, $\hs_\Ip = s_\Ip$. 

Now $N_e$ can be regarded as functional of $\hs_I(\sigma)$, 
\begin{equation}
N_e[\hs] = \int  \frac{ \ud \sigma \, 3H^2}{b + \hs_\sigma(\sigma)} + \lambda_d \int \ud\sigma \,  \hat{S}(\sigma) 
+ \sum_\Ip \int \frac{ \ud \sigma \, \lambda_\Ip \hs_{\Ip}(\sigma)}{b + \hs_\sigma(\sigma)} 
\end{equation}
where, following the constraint (\ref{d_cstr}), all the integrals are over the range $[0, \sigma_f]$. We have employed the Lagrangian multipliers $\{ \lambda_\Ip, \lambda_d\}$ to impose the constraints (\ref{i_cstr}) and (\ref{VScstr}). Using the variational method to find the stationary solution, 
we first require $\delta N_e/ \delta \lambda_\Ip = \delta N_e/\delta \hs_\Ip = 0$, so that
\begin{eqnarray}
\int \ud \sigma\, \frac{\hs_{\Ip}}{b + \hs_\sigma} = 0, \quad \lambda_\Ip = 2 \lambda_d \hs_\Ip ~. 
\end{eqnarray}
We get the solutions $\hs_\Ip(\sigma) = 0$, $\lambda_\Ip = 0$ for $\lambda_d \ne 0$. This means turning off transverse motions decreases the e-folds of inflation, i.e., for an arbitrary path along the $\sigma$ direction, any detour in the transverse directions always increases the travel time. In fact, the increase is unbounded from above. 

If we further set $\delta N_e/ \delta \lambda_d = \delta N_e/ \delta \hs_\sigma =0$, we have
\begin{eqnarray}
\frac{3H^2}{(b+\hs_\sigma)^2} = \lambda_d ~, \quad \int\ud\sigma\, \hs_\sigma = 0 ~,
\end{eqnarray}
which gives the solution $\hs_\sigma(\sigma) = 0$ and $\lambda_d = 3H^2/b^2$. 
This stationary solution is the minimum of the functional $N_e[\hs]$, 
that is, $N_e[0] = 3H^2\sigma_f/b$, which corresponds to turning off all the kinks in the potential. We therefore conclude that whenever $\hs(\sigma) \ne 0$, the number of e-folds always increases, that is, any detour from $V_c(\vphi)$ (\ref{bgV}) increases the e-folds of inflation. Again, the increase is unbounded from above. 

To understand the mechanism better, let us give an alternative proof. We introduce
\begin{equation} \label{bs}
\bs_\sigma \equiv - \frac{1}{\sigma_f} \int \hs_\sigma \,\ud\sigma = \frac{1}{\sigma_f} \sum_\Ip \int \ud\sigma\; \frac{\hs^2_\Ip}{b + \hs_\sigma} \ge 0~. 
\end{equation}
which simply follows from the constraint (\ref{VScstr}). Here $\bs_\sigma = 0$ if and only if $d=1$ or $\hs_\Ip = 0$. With transverse motion, $\bs_\sigma>0$. Since $\sigma$ is locally monotonic by construction, $b+\hs_\sigma(\sigma) > 0$ in the range $0 \le \sigma \le \sigma_f$, and we have $0 < b- \bs_\sigma \le b$. Now we consider the increase in e-folds due to the kinks (with $N_e[0] = 3H^2\sigma_f/b$),
\begin{eqnarray}
\label{Neinc}
\frac{N_e[\hs] - N_e[0]}{N_e[0]} = \frac{b}{\sigma_f} \int \frac{\ud \sigma}{b+\hs_\sigma(\sigma)} -1 
=  \frac{\bs_\sigma}{b-\bs_\sigma} + \left(\frac{b}{\sigma_f} \int \frac{\ud \sigma}{b+\hs_\sigma(\sigma)} - \frac{b}{b - \bs_\sigma} \right)  \ge 0 \nonumber 
\end{eqnarray}
where we have introduced an intermediate term containing $\bs_\sigma \ge 0$. Actually, both terms are semi-positive.
The first term in Eq.(\ref{Neinc}) is always positive with transverse motion.
In the presence of non-zero $\hs_I(\sigma)$, positivity of the second term follows from the inequality between the harmonic mean and the arithmetic mean, i.e., for $x_i > 0$ and $\sum \omega_i =1$, 
$\sum  \omega_i/{x_i} \ge \left( \sum \omega_i x_i \right)^{-1}$
(with the equality reached if and only if all the $x_i$'s are equal), where $\omega_i \rightarrow d\sigma/\sigma_f$ and   $x_i \rightarrow b+\hs_\sigma$.
We therefore see explicitly from Eq.(\ref{Neinc}) that $N_e[\hs] \ge N_e[0]$. For $d=1$, or no transverse motion, the increase is from the second term in Eq.(\ref{Neinc}) alone. For $d > 1$, if there is transverse motion, the first term always increases the e-folds of inflation.

To get an intuitive sense of the picture, consider Fig.\ref{fig_pot} for the $d=1$ case when $\veff(\sigma)=V(\phi)$. Potential $^\#1$ is the piecewise linear part of $V_c(\phi)$ from $\phi=0$ to $\phi=\sigma_f$. Potentials $^\#3$ ($K=3$) and $^\#4$ ($K=4$) can be considered as successively finer-grained approximations of the actual potential $^\#5$, i.e., we can treat $^\#3$ to be a coarse-grained potential with $2$ linear segments and $^\#4$ has a kink (with $K=3$) for each of the 2 segments of $^\#3$. For an arbitrary curve, we can break it into $K=3$ pieces, and the more we break it, the better it approximates the actual potential. More fine-grained features will always increase e-folds; here we have $N_e(^\#5) > N_e(^\#4) > N_e(^\#3) > N_e(^\#1)$. 
For $d>1$ with transverse motions, the curves in Fig.\ref{fig_pot} now represent $\veff(\sigma)$ (\ref{radpot1}). The \emph{actual} $d$-dimensional potential represented by $\veff(\sigma)$ always yields more e-folds than the corresponding $d=1$ potential shown, due to the detours of the inflaton field.

Most e-folds come from the regions with the smallest slope, where the constant $H$ and the slow-roll approximations are most valid. If one relaxes the $H \approx \mathrm{const}$ and the slow-roll approximations, one can show using the Hamilton-Jacobi approach that the solution corresponding to the minimal e-folds is given by an exponential potential. That is, for a given drop $\Delta H$ in the Hubble parameter over a given field range $[0, \phi_f]$, the minimal e-fold path is given by ($N_e =  \int \ud H/{\dot\phi} = - \frac{1}{2 {\mpl}^2} \int \ud \phi H/H'$)
\begin{eqnarray}
H(\phi)=H_i \exp \left(-\sqrt{2\epsilon} \frac{\phi}{\mpl}\right)
\end{eqnarray}
 where inflation takes place when $\epsilon <1$. Any deviation from this path will yield more e-folds. The linear piecewise potential can be replaced piecewise by this form for $H$. Both this $H$ and the corresponding exponential form of the potential approach linear when the $\epsilon$ parameter is small and/or the field range is small. It is clear that an arbitrary detour from such a potential will again lead to an increase in e-folds. So the qualitative properties of the scenario reman intact.

We expect the string scale and/or the Hubble scale to provide the cut-off on how fine-grained the actual potential can be. With appropriate characteristics of the actual potential, the enhancement can be large enough to give 60 or more e-folds. 

We have assumed that the slope is always monotonic, $s_\sigma (\sigma) >0$. It is argued in Ref.\cite{Tye:2006tg,Sarangi:2007jb} that the percolation probability $p \rightarrow 1$ in a random potential for large $d$. This is intuitively reasonable, since the inflaton has more choices of directions to move classically down a random potential as $d$ is large.  For the cosmic landscape, we expect $d \gg 1$, so the assumption $s_\sigma (\sigma) >0$ is quite natural. For small $d$, percolation is not assured. For $d=1$, the system typically does not percolate, so the monotonic condition on the slope is imposed by hand. The scenarios we have in mind have $d \gg 1$. 

\section{The Primordial Power Spectrum}

With $d$ canonical scalar fields, the scalar perturbation $v_I(k,\tau)$ evolves in conformal time $\tau$ according to 
\begin{equation}\label{mode}
v_I'' + \left(k^2 - \frac{a''}{a}\right) v_I + \left[a^2 V_{IJ} 
- \frac{1}{a^2} \left( \frac{a}{H} \phi'_I \phi'_J \right)' \right] v^J = 0 ~.
\end{equation}
If we model leading order effect due to the inflaton scattering by turning on $V_{IJ} \equiv \pd^2 V/\pd\phi_I \pd \phi_J$ at $t_k$, i.e.,
\begin{equation}\label{deltaij}
V_{IJ} = 3 H \Delta_{IJ} \delta(t-t_k)
\end{equation}
to leading order $\delta_I(k) = \Delta^{(k)}_{IJ} b^J$. The actual features in the potential may be more complicated with higher order derivatives in $V(\vphi)$. However, for two point function, only $V_{IJ}$ is relevant. Accordingly, $a^2 V_{IJ} = 3 a H \sum_i^K \Delta^{(i)}_{IJ} \delta(\tau - \tau_i)$, and the perturbation mode equation becomes (after ignoring $\epsilon$ terms)
\begin{equation}
\vv'' + \left(k^2 - \frac{a''}{a} \right)\vv - 3a H \sum_{i=1}^K \da^{(i)} \,\delta(\tau-\tau_i) \vv = 0  ~.
\end{equation}
with $\vv \equiv (v_1, v_2, \cdots v_d)^T$. Between the $i$-th and the $(i+1)$-th scattering, i.e. $\tau_i \le \tau \le \tau_{i+1}$, we can write
\begin{eqnarray}
\vv(k,\tau) = \valpha^i v^-(k,\tau) + \vbeta^i v^+(k,\tau) ~, \quad v^\pm (k,\tau) = \frac{1}{\sqrt{2k}} e^{\pm ik \tau} \left( 1 \pm \frac{i}{k\tau}  \right) 
\end{eqnarray}
Generalizing the approach in Ref.\cite{Starobinsky:1992ts}, we derive the coefficients $(\valpha^i, \vbeta^i)$ recursively, by matching the boundary conditions at each $\tau_i$. Starting with the Bunch-Davies vacuum $(\alpha^0_I,\beta^0_I)=(1,0)$ at $\tau = -\infty$, we get (to leading order in $\Delta^{(j)}$), 
\begin{eqnarray}
\alpha^n_I = 1 + \sum_{j=1}^n \sum_{J=1}^d \frac{3i}{2}\, {\Delta^{(j)}}^J_I \, (x_j^{-1} + x_j^{-3}) ~, \quad 
\beta^n_I = -\sum_{j=1}^n \sum_{J=1}^d \frac{3i}{2}\, {\Delta^{(j)}}^J_I  \, e^{2i x_j} (i + x_j)^2 ~. 
\label{coab}
\end{eqnarray}

Assuming $\la v^I v^J\ra \propto \delta^{IJ}$, the adiabatic power spectrum $P_\zeta$ is
\begin{eqnarray}
&& P_\zeta(k) = \frac{H^4}{4\pi^2 \dot\phi^2} \sum_{I=1}^d \frac{|\beta_I - \alpha_I|^2}{d} \Big|_{\tau\to 0}  
= \frac{H^4 }{4\pi^2 {\dot\phi}^2} \left( 1 + \sum_{i=1}^K  B^i p_1(x_i) \right) \label{pk} ~, \\
&& B^i \equiv \frac{1}{d} \sum_{I,J} \Delta^{(i)}_{IJ} ~, \quad 
p_1(x_i) = 3(x_i^{-3} - x_i^{-1})\sin{2x_i} - 6 x_i^{-2} \cos{2x_i} ~.
\end{eqnarray}
Incorporating the time dependence of $H$, we can parametrize 
$H^4/(4\pi^2 {\dot\phi}^2) = A_s \left(k/k_0\right)^{n_s - 1}$, 
which represents the gross feature of the power spectrum, with $A_s = 2.41 \times 10^{-9}$, $n_s = 0.95$ at $k_0 = 0.002 \mpc^{-1}$ \cite{Komatsu:2008hk}. The inflaton scatterings will impart fluctuations of magnitude $B^i$ on such a smooth power spectrum (Fig.\ref{fig_pk}). For single field $d=1$, the features correspond to sudden changes in the potential gradient, while for multifield $d \gg 1$, the features arise due to the turning of inflaton trajectory. The turning angle can be quite large $\Delta_{IJ} \sim 1$, and still be compatible with data since $B^i \sim \Delta_{IJ}/d$ for the multifield case. Although features arise due to different reasons in the single field and multifield scenario, one can not tell the difference from their appearance in the power spectrum.

If we assume that a few of such features already have been detected in the  data \cite{Nolta:2008ih,Nagata:2008tk,Nagata:2008zj,Nicholson:2009pi}, 
two possible scenarios seem to emerge: (1) the spacings between fluctuations (i.e., scatterings) can be many units of $l$, say $\Delta l \sim 100$; (2) although the spacings between fluctuations are too small to be resolved, a few occasional fluctuations are big enough and relatively well spaced to be observed .

Fluctuations in the primordial power spectrum may have a chance to show up in the CMB temperature power spectrum. Looking at the un-binned TT CMB power spectrum released by WMAP \cite{Nolta:2008ih}, one sees that it fluctuates for about $10\%$ (for $100 \lesssim l \lesssim 800$). However, due to the large error bars of the same magnitude ($10 \%$), it is hard to claim any features based on current data. The usual data analysis approach bins a few tens of multiple moments together to give a relatively smooth TT power spectrum over all angular scales. However, given the presence of fluctuations in the primordial power spectrum, the data binning approach will have smoothed out these features if they appear frequent enough. 

PLANCK can provide cosmic variance limited measurements up to $l\sim 2000$. In fact $\delta C^{TT}_l/C^{TT}_l \lesssim 5\%$ for PLANCK in the range $500 \lesssim l \lesssim 2000$. So each $10\%$ fluctuation in $C_l^{TT}$ will be a $2\sigma$ effect for PLANCK. Once such features are revealed in the TT power spectrum, they must appear in the TE/EE  polarization power spectra  at the same angular scales, providing an important check on our predictions. 

The presence of fluctuations in the power spectrum will imply that the inflaton is mobile in the random potential, i.e., the cosmic landscape. Since our universe has gone through the cosmic landscape only once in this scenario, the fluctuations in the CMB power spectrum will reveal this particular path of our past history. 

\section{Discussions}

In the paper, we have presented a scenario where a large number of scalar fields collectively contribute to cosmic inflation. It is important to distinguish our scenario from existing multifield scenarios like assisted inflation \cite{Liddle:1998jc} and N-flation \cite{Dimopoulos:2005ac, Easther:2005zr}.

In the assisted inflation scenario, different fields have identical uncoupled potential. The trajectory of inflaton field is most likely a smooth path close to a straight line during the few e-folds relevant for CMB observation, which does not allow entropic perturbations to play any role. If the single field potential has some features like steps or kinks, each field will experience the same feature as it rolls down the potential. This will create identical features on the power spectrum with the same shape/magnitude.

In the N-flation scenario, a large number of axion fields play the role of the inflatons. If we only keep the leading order one-instanton terms, and Taylor expand each axion potential around its minimum, we get a set of fields with different masses. At one-instanton level, different inflaton fields still decouple, but since their masses are all different, the inflaton trajectory could be curved. However, the turning of the inflaton field is very mild in the slow-roll regime, and again does not lead to sharp features in the power spectrum. One can keep the sub-leading multi-instanton terms in the potential; this will explicitly introduce couplings between the fields, and could make the inflaton path more convoluted, therefore providing an explicit realization of the meandering inflaton in string theory. In this case, our approach provides a different perspective to the analysis of these complicated scenarios. In particular, our analysis implies that the simplified analysis carried out in the literature on such a scenario
may substantially under-estimate the e-folds produced by the model. 

There has been a lot of conjectures regarding the features in the CMB power spectrum. However, it is not clear what is the origin of such sharp features from string theory. In Ref.\cite{Bean:2008na} and \cite{Flauger:2009ab}, features in the potential are well motivated to arise from duality cascade or monodromy. Here we provide another natural motivation: the meandering of the inflaton in the cosmic landscape. Regarding observations, duality cascade and monodromy lead to features on the power spectrum with predicted location and magnitude, while here we have random fluctuations in the power spectrum which is extremely hard to be picked out by current data analysis approaches.

\vspace{0.3cm}
{\bf Acknowledgments:}
We thank Rachel Bean, Scott Dodelson, Liam McAllister, David Spergel, Jun'ichi Yokoyama, Yang Zhang for valuable discussions. This work is supported by the National Science Foundation under grant PHY-0355005.

\end{document}